\documentclass[reprint,aps,prl]{revtex4-2} 

\usepackage{graphicx}
\usepackage{dcolumn}
\usepackage{bm}
\usepackage{amsmath}
\usepackage{xcolor}

\begin{document}

\preprint{APS/123-QED}

\title{Perfect Absorption in the Critically Damped Regime}

\author{Viacheslav V. Medvedev}
\email{medvedev@phystech.edu}
\affiliation{Institute of Spectroscopy, Russian Academy of Sciences, Troitsk, Moscow 108840, Russia}

\begin{abstract}
We revisit the conditions for perfect electromagnetic absorption in a homogeneous lossy layer on a reflecting substrate. Using phasor diagram analysis, we demonstrate a fundamental physical similarity between the high-refractive-index limit (classic Dallenbach quarter-wavelength absorbers) and the epsilon-near-zero regime (half-wavelength resonances). In both extremes, perfect absorption relies on long cyclic multipath propagation and gradual amplitude decay. Crucially, we uncover that in the intermediate regime near $n = 1$, this picture changes fundamentally: the trapping efficiency drastically increases, and backscattering is eliminated almost instantaneously within a single round-trip. Using temporal coupled-mode theory, we prove that this low-contrast state minimizes the system's quality factor to a global minimum of $Q \approx 0.69$. This critically damped state mirrors universal highly damped stabilization principles found in acoustics and mechanics, driving anomalous spectral broadening and enabling nearly instantaneous dissipation of ultrashort pulses without time-domain ringing.
\end{abstract}

\maketitle


Controlling subwavelength electromagnetic absorption is a cornerstone challenge in photonics and radiophysics, driving innovations in energy harvesting, sensing, and stealth technologies. At the heart of these wave-trapping phenomena lies destructive interference, the ultimate mechanism for maximizing energy attenuation. The simplest configuration for such trapping is the Dallenbach absorber \cite{dallenbach1938reflection,Tretyakov2016IEEE,kotsuka2019electromagnetic}, which comprises a single lossy dielectric layer backed by a perfectly reflecting metallic substrate. In its original concept, perfect absorption is achieved when the layer’s optical thickness equals odd multiples of a quarter-wavelength ($\lambda_0/4$), triggering destructive interference that cancels the primary reflection. While conceptually similar to anti-reflection coatings, the Dallenbach absorber blocks transmission via the metallic backing, trapping and dissipating the wave within the lossy medium. Recently, this paradigm experienced a major revival in the optical regime following the work by Kats \textit{et al.} \cite{Kats2013NatureMaterials}, who demonstrated that nontrivial phase shifts at lossy-film/metal interfaces can drastically reduce the required layer thickness well below the conventional $\lambda_0/4$ limit.

This renewed interest has triggered rigorous theoretical efforts to map the exact conditions for perfect absorption \cite{Park2014ACSPhotonics,Park2015OL,Lezaun2024PRB,Medvedev2021IEEE,Medvedev2022OM,johns2022tailoring}. Notably, Park \textit{et al.} \cite{Park2015OL} re-evaluated the perfect electrical conductor (PEC) substrate scenario, showing that total absorption within a film of refractive index $n$ requires an extinction coefficient $\kappa = 2/\pi$ at an optical thickness $nd = \lambda_0/4$. Subsequently, Medvedev \cite{Medvedev2022OM} extended this framework to epsilon-near-zero (ENZ) media, revealing that the optimal thickness shifts to a half-wavelength resonance while the extinction coefficient follows a distinct scaling law: $\kappa = n^2 / \pi$. Despite the wealth of literature \cite{Pottel1959ZPhysik,Fernandez1985ElectronLett,Valenzuela1996IEEE,Rozanov2000IEEE,Park2014ACSPhotonics,Park2015OL,Lezaun2024PRB,Medvedev2021IEEE,Medvedev2022OM,johns2022tailoring}, a unified physical picture detailing how the absorption regime continuously transforms from the half-wavelength ENZ limit to the classical Dallenbach solution remains absent.

In this Work, we address this gap by comprehensively analyzing total absorption conditions using a dual-method approach that combines partial-wave phasor diagrams with temporal coupled-mode theory (TCMT). We demonstrate a fundamental physical similarity between the classical Dallenbach and ENZ regimes: both operate as high-$Q$ resonators sustained by a long train of multiple internal reflections. Crucially, however, we discover a distinct, impedance-matched absorption regime near $n = 1$, where the wave enters the film with minimal reflection and attenuates within a single round trip. TCMT reveals that the quality factor drops to its absolute theoretical minimum of $Q \approx 0.69$. This extreme state marks the transition into a critically damped oscillator regime, forcing the incident energy to dissipate within nearly a single optical cycle without time-domain ringing, thereby unlocking new pathways for absorbing ultrashort pulses. We show that this critically damped state is uniquely robust against phase detuning, providing the underlying mechanism for anomalous spectral broadening. Remarkably, this operating principle mirrors universal, highly damped stabilization mechanisms engineered in high-performance acoustic dampers and automotive suspensions, as well as those found in biological locomotor systems.


\begin{figure*}[t]
\centering
\includegraphics[width=0.7\textwidth]{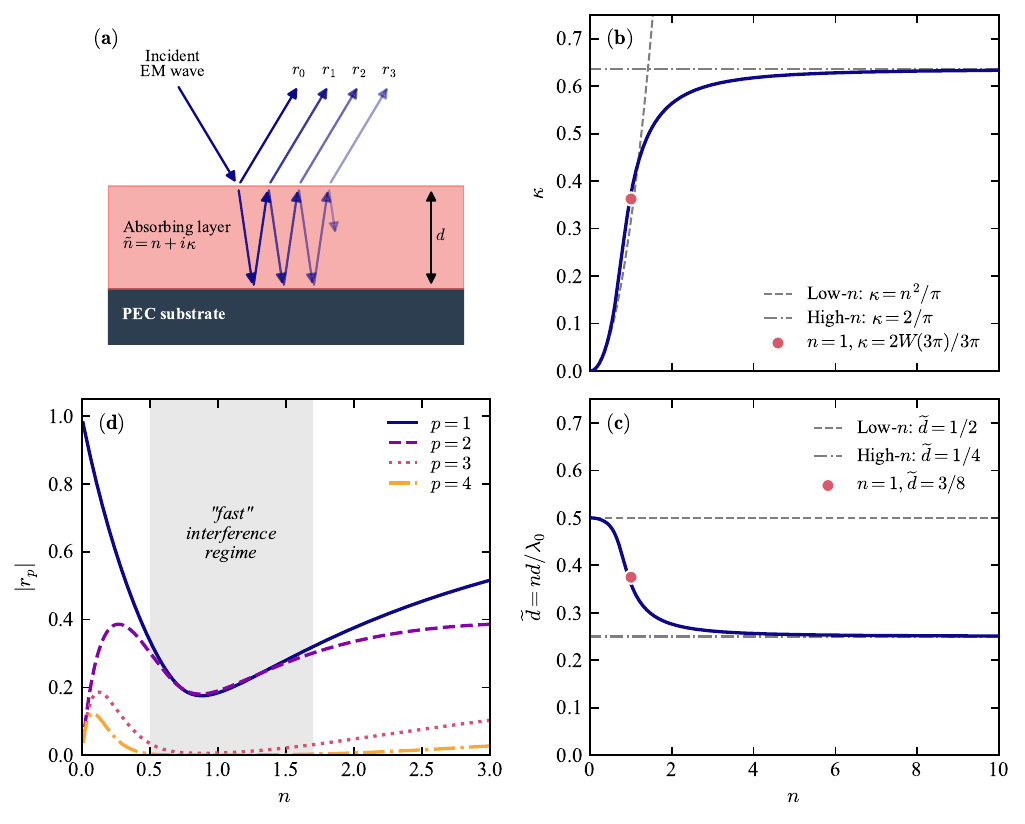}
\caption{(a) Schematic of a homogeneous lossy layer ($l$) of thickness $d$ and complex refractive index $\widetilde{n} = n + i\kappa$ on a perfect electric conductor (PEC) substrate ($s$). Arrows indicate the incident wave from air ($a$, $n=1$) and the multiwave propagation of constituent partial waves. Calculated layer parameters for perfect absorption ($r=0$, fundamental mode $m=0$) versus the real refractive index $n$: (b) optimal extinction coefficient $\kappa$ and (c) normalized optical thickness $n d/\lambda_0$. Solid lines denote the exact numerical solution; circular markers show the approximate analytical solution at $n=1$. Dashed gray lines indicate analytical asymptotes for $n \ll 1$ \cite{Medvedev2022OM}, and dash-dotted lines mark the classical Dallenbach limit for $n \gg 1$ \cite{dallenbach1938reflection}. (d) Amplitudes of the first three partial waves $|r_1|$, $|r_2|$, and $|r_3|$ versus $n$ under perfect absorption. The shaded gray area outlines the low-contrast, fast-interference window ($0.5 < n < 1.7$).}
\label{fig_system_and_parameters}
\end{figure*}
Consider a monochromatic plane wave normally incident from air ($n_a=1$) onto a homogeneous dissipative layer of thickness $d$ backed by a PEC substrate [Fig.~\ref{fig_system_and_parameters}(a)]. At the vacuum wavelength $\lambda_0$ (angular frequency $\omega_0 = 2\pi c/\lambda_0$), the layer has a complex refractive index $n + i\kappa$, yielding the wave number $k = 2\pi(n + i\kappa)/\lambda_0$. We define the normalized optical thickness as $\tilde{d} = nd/\lambda_0$. Since the PEC substrate is perfectly opaque, energy conservation simplifies to $A = 1 - R$. Perfect absorption ($A = 1$) requires eliminating the power reflection coefficient ($R = |r|^2 = 0$). Using the Airy formalism \cite{BornWolf}, we expand the total amplitude reflection coefficient $r$ into a series of constituent partial waves $r_{p}$:
\begin{equation}
r = \sum_{p=1}^{\infty} r_{p}, \quad  
r_{p} = \begin{cases} 
r_{al}, & p = 1, \\
\frac{r_{al}^2 - 1}{r_{al}} \left( r_{al} \exp(2 i k d) \right)^{p-1}, & p > 1,
\end{cases}
\label{eq_reflection_Airy}
\end{equation}
where $r_{al} = (1 - n - i\kappa)/(1 + n + i\kappa)$ is the Fresnel reflection coefficient at the air-layer interface. Summing this geometric progression yields the total reflection coefficient:
\begin{equation}
    r = \frac{r_{al} - \exp(2 i k d)}{1 - r_{al}\exp(2 i k d)}.
    \label{eq_reflection_reduced}
\end{equation} 
The zero-reflection conditions across the $(n, \kappa, \tilde{d})$ parametric space are mapped by numerically solving $r_{al} - \exp(2 i k d) = 0$~\cite{Medvedev2022OM}, as illustrated in Figs.~\ref{fig_system_and_parameters}(b) and \ref{fig_system_and_parameters}(c). In the high-index regime ($n > 2$), the solutions approach the classic Dallenbach limit: $\tilde{d} \to 1/4$ (quarter-wavelength layer) and $\kappa \to 2/\pi$~\cite{Tretyakov2016IEEE}. Conversely, in the ENZ limit ($n \to 0$), they follow the half-wavelength resonance scaling: $\tilde{d} \to 1/2$ and $\kappa \to n^2/\pi$~\cite{Medvedev2022OM}. While $\kappa$ and $\tilde{d}$ vary monotonically between these asymptotes, the key physical question is how the underlying interference mechanism evolves across this $r=0$ solution space.

\begin{figure*}[t]
\centering
\includegraphics[width=0.8\linewidth]{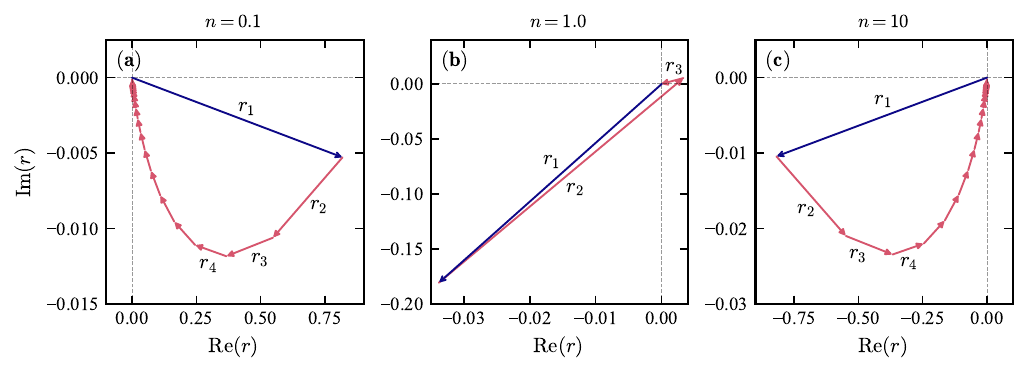}
\caption{Phasor diagrams of the complex partial-wave amplitudes $r_p$ contributing to perfect absorption for (a) $n = 0.1$ (ENZ limit), (b) $n = 1.0$ (low-contrast rapid-interference regime), and (c) $n = 10.0$ (classical Dallenbach limit). Spatial scales across panels (a)--(c) are individually adjusted for visual clarity. The first few partial wave vectors $r_p$ are explicitly labeled.}
\label{fig_partial_waves}
\end{figure*}

To uncover the underlying wave dynamics, we analyze phasor diagrams of the constituent partial waves $r_p$ [Eq.~\eqref{eq_reflection_Airy}] for three representative cases: $n=10$, $0.1$, and $1$ (Fig.~\ref{fig_partial_waves}). In the high-index Dallenbach limit ($n=10$), high optical contrast produces a strong primary reflection, $r_1 \approx 0.82 \exp(i \pi)$, which is canceled by a long, slowly decaying sequence of subsequent multipath reflections [Fig.~\ref{fig_partial_waves}(c)]. A qualitatively similar behavior occurs in the extreme ENZ limit ($n=0.1$), where the first partial wave vector rotates to $r_1 \approx 0.82 \exp(2 i \pi)$ [Fig.~\ref{fig_partial_waves}(a)]. This $\pi$-phase shift stems from the sign inversion of the real part of the Fresnel coefficient $r_{al}$ as $n$ drops below unity. Crucially, this reversal alters the global phase-matching condition, explaining why destructive interference requires a quarter-wavelength thickness ($\tilde{d} \to 1/4$) in the Dallenbach limit but scales to a half-wavelength resonance ($\tilde{d} \to 1/2$) in the ENZ regime.

Conversely, in the intermediate low-contrast regime ($n=1$), the physical mechanism changes fundamentally. As shown in Fig.~\ref{fig_partial_waves}(b), the primary reflection is significantly weaker ($r_1 \approx 0.18 \exp(1.44 i \pi)$), allowing perfect destructive interference to be established by just two subsequent partial waves. Instead of relying on an extended coherent series, the energy here is dissipated within nearly a single round-trip. Numerical mapping of the partial amplitudes $|r_p|$ versus $n$ under the perfect absorption condition [Fig.~\ref{fig_system_and_parameters}(d)] confirms that this rapid compensation mechanism remains robust across a broad design window ($0.5 < n < 1.7$), where the third contribution $|r_3|$ remains below $10\%$ and the fourth $|r_4|$ drops to just $1\%$ of $|r_1|$.

Notably, at the point $n=1$, which symbolizes this newly uncovered regime of rapid interference, an approximate analytical solution can be derived via the Lambert $W$ function: $d = 3\lambda_0/8$ and $\kappa = \frac{2}{3\pi} W(3\pi) \approx 0.36$ (see Sec. I of the Supplemental Material). This analytical solution is highlighted with circular markers in Figs.~\ref{fig_system_and_parameters}(b) and \ref{fig_system_and_parameters}(c).


\begin{figure*}[t]
\centering
\includegraphics[width=0.7\textwidth]{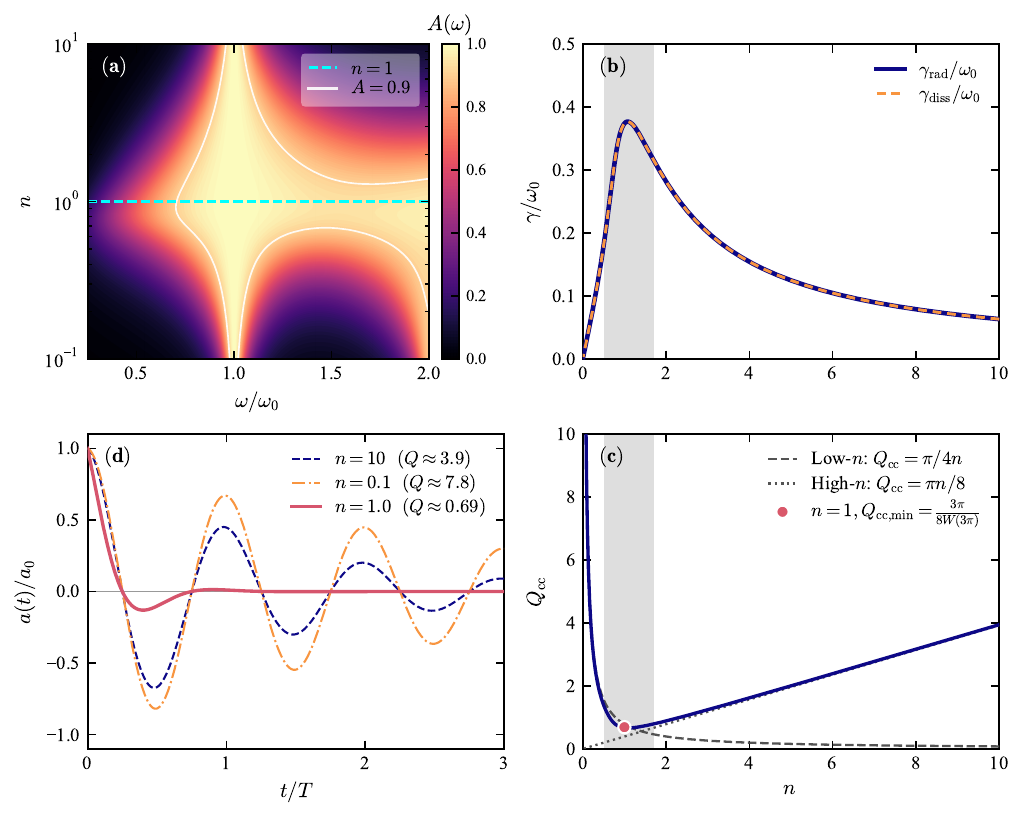}
\caption{Spectral and temporal dynamics of the optimized system: (a) Spectral absorption landscape $A(\omega, n)$ under the dispersionless approximation ($n, \kappa = \text{const}$). (b) TCMT radiative decay rate $\gamma_{\text{rad}}$ (solid line) and dissipative decay rate $\gamma_{\text{diss}}$ (dashed line) normalized to the central frequency $\omega_0$ versus $n$. Both rates balance identically, exhibiting a pronounced maximum near $n = 1$. The shaded gray area outlines the fast-interference window ($0.5 < n < 1.7$). (c) Critical coupling quality factor $Q_{\text{cc}}$ of the absorbing mode versus $n$. The circular marker indicates the approximate analytical minimum value $Q_{\text{cc, min}} \approx 0.69$ at $n=1$; dashed and dotted gray lines show the near-zero-index ($n \ll 1$) and high-contrast ($n \gg 1$) limiting scalings, respectively. (d) Temporal evolution of the normalized mode amplitude $a(t)/a_0$ as a function of time $t$ normalized to the optical oscillation period $T$ for $n=0.1$ (dash-dotted line), $n=1.0$ (solid line), and $n=10.0$ (dashed line), demonstrating that the field is critically damped within a single optical cycle exclusively in the low-contrast regime ($n=1$).}
\label{fig_tcmt_and_spectra}
\end{figure*}

This rapid-damping mechanism fundamentally alters the spectral response of the system. Assuming a dispersionless medium ($n, \kappa = \text{const}$), Fig.~\ref{fig_tcmt_and_spectra}(a) maps the absorption landscape in the $(\omega/\omega_0, n)$ plane. While the high- and low-contrast regimes exhibit narrow, highly dispersive absorption peaks, the vicinity of $n \approx 1$ displays an anomalous spectral broadening. The interference condition becomes so resilient to frequency detuning that neighboring absorption modes overlap, forming a continuous plateau with absorption $A > 90\%$ across the entire inter-resonance interval. Importantly, this low sensitivity to phase detuning implies that the system is inherently robust against fabrication tolerances in thickness $d$ and thermal fluctuations of the refractive index.

To uncover the dynamic nature of this absorption mechanism, we transition from the interference-based picture to the temporal coupled-mode theory (TCMT) framework~\cite{Haus1984, Fan2003JOSA}. The layered system acts as a single-port open resonator, whose near-resonance dynamics are governed by radiative ($\gamma_{\text{rad}}$) and dissipative ($\gamma_{\text{diss}}$) decay rates (see Sec.~II of the Supplemental Material for details; similar absorbing topologies are discussed in~\cite{Piper2014ACSPhoton}). Within this framework, perfect absorption strictly corresponds to the critical coupling regime, where the decay channels balance identically: $\gamma_{\text{rad}} = \gamma_{\text{diss}}$.

To map these phenomenological TCMT parameters onto the layer's physical properties, we employ the thin-film approximations $\gamma_{\text{rad}} \approx \frac{c}{2nd} \ln (1/|r_{al}|)$ and $\gamma_{\text{diss}} \approx \frac{\omega_0 \kappa}{n}$ (see Sec.~II of the Supplemental Material). Evaluating these rates using the optimal parameters $(n, \kappa, \tilde{d})$ confirms that $\gamma_{\text{rad}} = \gamma_{\text{diss}}$ holds across all $n$ [Fig.~\ref{fig_tcmt_and_spectra}(b)], ensuring critical coupling. Crucially, both decay rates exhibit a pronounced maximum near $n = 1$.

The quality factor $Q$ of an open resonator is defined through the total energy decay rate (see Sec.~II of the Supplemental Material), which under critical coupling ($\gamma_{\text{rad}} = \gamma_{\text{diss}}$) simplifies to:\begin{equation}Q_{\text{cc}}(n) = \frac{\omega_0}{4\gamma_{\text{diss}}} = \frac{n}{4\kappa(n)}.\label{eq_Q_cc_min_form}\end{equation}Using the asymptotic scalings for $\kappa(n)$ (see Sec.~II of the Supplemental Material), we obtain the limiting behaviors: $Q_{\text{cc}} \to \pi/(4n)$ for $n \ll 1$ and $Q_{\text{cc}} \to \pi n/8$ for $n \gg 1$ [dashed and dotted lines in Fig.~\ref{fig_tcmt_and_spectra}(c)]. The quality factor reaches a global minimum near $n = 1$, which evaluates analytically via the Lambert $W$ function as:
\begin{equation}
Q_{\text{cc, min}} = \frac{3\pi}{8 W(3\pi)} \approx 0.69,
\label{eq_Q_cc_min}
\end{equation}
as highlighted by the circular marker in Fig.~\ref{fig_tcmt_and_spectra}(c).

Furthermore, Fig.~\ref{fig_tcmt_and_spectra}(d) illustrates the decaying transient oscillations of the cavity field amplitude $a(t)/a_0$ calculated within the TCMT framework (see Sec. III of the Supplemental Material). In the case of $n=1$, the internal field within the layer decays almost entirely within a single optical cycle of the incident light, translating to an energy suppression of approximately four orders of magnitude (${\sim}10^{-4}$) per period. Note that although the TCMT formalism traditionally assumes relatively long-lived states ($Q \gg 1$), the perfect algebraic agreement with the exact Airy equations demonstrates that it provides an exceptionally accurate phenomenological mapping for our system, successfully extending its predictive power deep into the sub-unity regime ($Q < 1$).


\textit{Material platforms and physical realism.}---Practically, the low-contrast regime occupies a design window bounded by $0.5 < n < 1.7$ and $0.1 < \kappa < 0.5$. Optical database analysis~\cite{Polyanskiy2024SciData} reveals that such constitutive parameters are naturally available in polar dielectrics near phonon resonances (mid- and far-infrared)~\cite{Franta2015AO, Kischkat2012AO, Franta2015ProcSPIE, Herguedas2023Nanomaterials} and in conducting metal oxides near plasmon resonances (near-infrared)~\cite{Cleary2015OMEx, Medvedev2022OM}.

However, strong material dispersion inherent to these resonances inevitably narrows the absorption peaks compared to the ideal dispersionless case. To engineer true ultra-broadband absorbers, one must develop artificial metamaterials that provide the target $n$ and $\kappa$ while suppressing dispersion across the operational band. A promising strategy involves porous composite media utilizing a low-index matrix (e.g., fluorides) doped with metallic nanoparticles or carbon nanotubes; tailoring the inclusion geometry can position scattering resonances far from the desired spectral window to ensure dispersionless attenuation.


\textit{Interdisciplinary analogies.}---The global minimum quality factor $Q \approx 0.69$ reveals that optimal wave attenuation in planar layers is governed by the same universal stabilization principles that evolution and engineering have independently converged upon across acoustics, mechanics, and biology. This subunity threshold fundamentally mirrors the ``maximally flat'' Butterworth response in acoustics ($Q \approx 0.707$)~\cite{Small1972}, optimized ride-handling damping ratios in automotive suspensions ($0.6 < Q < 1.6$)~\cite{Gillespie1992, Milliken1995}, and the near-critical damping of biological musculoskeletal systems optimized to instantly dissipate inertial energy and eliminate tissue-damaging secondary oscillations~\cite{Piovesan2013, Alexander1988}. These diverse systems utilize highly damped states because they represent the strict mathematical boundary ensuring the fastest return to equilibrium. In photonics, this boundary unlocks distortionless management of single- and sub-cycle optical pulses. Unlike conventional high-$Q$ Dallenbach or ENZ absorbers, which induce prolonged time-domain ringing due to slow multipath interference build-up, this low-contrast lumped damper eliminates temporal reflection artifacts almost instantaneously. By achieving full energy dissipation within just two to three partial-wave cycles, the system preserves the pristine temporal envelope and phase profiles of ultrashort wave packets, offering a critical design paradigm for attosecond physics and ultrafast signal processing.


In conclusion, we have uncovered a low-contrast absorption regime in lossy layers on reflecting substrates that redefines the boundaries of interference-based wave trapping. By bridging the multiwave Airy formalism with temporal coupled-mode theory, we demonstrated that near $n = 1$, the system transitions into a critically damped state characterized by a global minimum quality factor $Q \approx 0.69$, acting as a lumped electromagnetic damper. This mechanism eliminates transient time-domain ringing and phase distortions, enabling nearly instantaneous dissipation of ultrashort pulses within two to three partial-wave cycles. Although natural polar dielectrics and conducting oxides provide the target $n$ and $\kappa$ parameters near their phonon and plasmon resonances, strong material dispersion narrows the operational bandwidth. To bypass this limitation, we proposed a design strategy based on artificial composites with reduced dispersion of optical constants.

\nocite{*}
\bibliography{apssamp}

\clearpage 
\onecolumngrid 

\setcounter{page}{1}       
\setcounter{equation}{0}   
\setcounter{figure}{0}     
\setcounter{table}{0}      
\setcounter{secnumdepth}{2} 

\renewcommand{\thepage}{S\arabic{page}} 
\renewcommand{\theequation}{S\arabic{equation}}
\renewcommand{\thefigure}{S\arabic{figure}}
\renewcommand{\thetable}{S\arabic{table}}
\renewcommand{\thesection}{\Roman{section}} 

\begin{center}
    \vspace*{0.5cm}
    \textbf{\large SUPPLEMENTARY MATERIALS} \\[0.2cm]
    \textbf{\large Perfect Absorption in the Critically Damped Regime} \\[0.4cm]
    V.V. Medvedev \\
    \textit{Institute of Spectroscopy, Russian Academy of Sciences, Troitsk, Moscow 108840, Russia}
    \vspace*{0.8cm}
\end{center}

\section{Analytical Solution for Perfect Absorption Conditions at $n=1$}
\label{sm_sec_n=1}

To establish the analytical reference for the low-contrast transition regime, we evaluate the perfect absorption condition, $r_{al} = \exp(2ikd)$, at the specific boundary where the real part of the refractive index equals unity ($n=1$). At this point, the complex refractive index simplifies to $\widetilde{n} = 1 + i\kappa$, and the wave number within the dissipative layer is defined as $k = \frac{\omega_0}{c}(1 + i\kappa)$.

Substituting $\widetilde{n}$ into the Fresnel reflection coefficient $r_{al} = (1-\widetilde{n})/(1+\widetilde{n})$ yields:
\begin{equation} 
r_{al} = \frac{-i\kappa}{2 + i\kappa}. 
\label{eq_supp_ral_n1} 
\end{equation}
Consequently, the governing equation for perfect absorption, $r_{al} - \exp(2 i k d) = 0$, can be explicitly rewritten as:
\begin{equation} 
\frac{-i\kappa}{2 + i\kappa} = \exp\left(2i \frac{\omega_0}{c} d\right) \exp\left(-2 \frac{\omega_0}{c} \kappa d\right). 
\label{eq_supp_trans_n1} 
\end{equation}

To determine the geometric scale of this transition state, we invoke the phase matching condition from the fundamental absorption mode ($m=0$). At $n=1$, the phase requirement couples the physical layer thickness to the interfacial reflection phase via $4\pi d / \lambda_0 = \arg(r_{al})$. From Eq.~\eqref{eq_supp_ral_n1}, for small to moderate values of the extinction coefficient $\kappa$, the reflection phase $\arg(r_{al})$ remains in the immediate vicinity of $1.5\pi$. Specifically, for the targeted optimized state where $\kappa \approx 0.37$, direct computation yields $\arg(r_{al}) \approx 1.51\pi \approx 3\pi/2$. Substituting this approximation into the phase condition determines the fundamental geometric scale of the layer:
\begin{equation} 
d = \frac{3}{8}\lambda_0. 
\label{eq_supp_3/8_result} 
\end{equation}

Next, we substitute the derived thickness $d$ back into the amplitude matching condition, $|r_{al}| = \exp(-4\pi\kappa d / \lambda_0)$. Utilizing the geometric scaling from Eq.~\eqref{eq_supp_3/8_result}, the amplitude condition takes the form:
\begin{equation} 
\frac{\kappa}{\sqrt{4 + \kappa^2}} = \exp\left(-\frac{3\pi}{2}\kappa\right). 
\label{eq_supp_amp_final} 
\end{equation}
By neglecting the small higher-order term $\kappa^2$ in the denominator ($\kappa^2 \ll 4$), Eq.~\eqref{eq_supp_amp_final} simplifies to $\frac{\kappa}{2} \approx \exp\left(-\frac{3\pi}{2}\kappa\right)$. Multiplying both sides by $3\pi \exp\left(\frac{3\pi}{2}\kappa\right)$ casts the expression into the transcendental form $x e^x = y$:
\begin{equation} 
\left(\frac{3\pi}{2}\kappa\right) \exp\left(\frac{3\pi}{2}\kappa\right) \approx 3\pi. 
\end{equation}
The analytical solution to this equation is uniquely expressed through the principal branch of the Lambert $W$ function:
\begin{equation} 
\kappa \approx \frac{2}{3\pi} W\left(3\pi\right) \approx 0.362. 
\label{eq_supp_kappa_W} 
\end{equation}
A rigorous numerical solution of the full, unapproximated system yields the exact values of $\kappa \approx 0.371$ and $d \approx 0.364\lambda_0$, validating the high fidelity of the $3/8$-wavelength analytical reference.

\section{Temporal Coupled-Mode Theory (TCMT) Framework}
\label{sec_supp_tcmt_full}

Following the established single-port temporal coupled-mode theory (TCMT) framework~\cite{Haus1984, Fan2003JOSA}, we model the lossy layer on a reflecting substrate as a localized resonant mode coupled to a single scattering port (the ambient continuum). Assuming a time-harmonic dependence of $\exp(-i\omega t)$, the dynamic evolution of the mode amplitude $a$ (normalized such that $|a|^2$ represents the stored energy) under excitation by an incident wave $s_{+}$ and the subsequent formation of the reflected wave $s_{-}$ are governed by the following coupled linear differential equations:
\begin{equation}
\frac{da}{dt} = (-i\omega_0 - \gamma_{\text{rad}} - \gamma_{\text{diss}})a + \chi s_{+},
\label{eq_supp_da_dt}
\end{equation}
\begin{equation}
s_{-} = C s_{+} + \chi a,
\label{eq_supp_s_minus}
\end{equation}
where $\omega_0$ is the central resonant frequency, $\gamma_{\text{rad}}$ is the radiative decay rate describing energy leakage back into the ambient vacuum, and $\gamma_{\text{diss}}$ is the internal dissipative decay rate due to material loss. 

Energy conservation and time-reversal symmetry constraints dictate a strict relationship between the coupling coefficient $\chi$ and the radiative decay rate, yielding $\chi = \sqrt{2\gamma_{\text{rad}}}$. Concurrently, the background scattering coefficient is given by $C = -1$, which physical mirrors the instantaneous $\pi$ phase shift experienced by a plane wave upon direct reflection from a perfect electric conductor (PEC) boundary.

To analyze the spectral response of the system, we transform Eqs.~\eqref{eq_supp_da_dt} and \eqref{eq_supp_s_minus} into the frequency domain by substituting $d/dt \to -i\omega$. Eliminating the internal mode amplitude $a$, we readily obtain the standard complex amplitude reflection coefficient $r(\omega) = s_{-}/s_{+}$:
\begin{equation} 
r(\omega) = \frac{i(\omega - \omega_0) + \gamma_{\text{rad}} - \gamma_{\text{diss}}}{-i(\omega - \omega_0) + \gamma_{\text{rad}} + \gamma_{\text{diss}}}. 
\label{eq_supp_r_tcmt} 
\end{equation}
The energetic absorption coefficient of this single-port system stands as $A(\omega) = 1 - |r(\omega)|^2$. At the exact resonance frequency ($\omega = \omega_0$), Eq.~\eqref{eq_supp_r_tcmt} simplifies to $r(\omega_0) = (\gamma_{\text{rad}} - \gamma_{\text{diss}})/(\gamma_{\text{rad}} + \gamma_{\text{diss}})$. Consequently, achieving perfect absorption ($A = 1$, or equivalently $r = 0$) strictly requires the perfect balance between the two energy extraction channels:
\begin{equation}
\gamma_{\text{rad}} = \gamma_{\text{diss}}.
\label{eq_supp_critical_coupling}
\end{equation}
Equation~\eqref{eq_supp_critical_coupling} defines the critical coupling condition for the absorber.

To bridge this phenomenological TCMT model with the macroscopic parameters of the homogeneous layer ($n, \kappa, d$), we employ a round-trip wave-packet approximation. A localized wave packet trapped inside a layer of thickness $d$ travels at a group velocity $v_g \approx c/n$. The characteristic round-trip time required for the energy to traverse the layer twice and return to the front interface is given by:
\begin{equation}
t_{\text{rt}} = \frac{2nd}{c}.
\label{eq_supp_trt}
\end{equation}
The radiative decay rate $\gamma_{\text{rad}}$ represents the fractional energy leakage per unit time through the air-layer interface into the ambient vacuum. This rate is governed by the energy reflection coefficient at the front boundary, $|r_{al}|^2$, where $r_{al} = (1 - \widetilde{n})/(1 + \widetilde{n})$ is the Fresnel reflection coefficient. Expressing the exponential decay of energy via the round-trip timeline yields:
\begin{equation}
\gamma_{\text{rad}} \approx \frac{1}{2} \frac{\ln(1/|r_{al}|^2)}{t_{\text{rt}}} = \frac{c}{2nd} \ln \left( \frac{1}{|r_{al}|} \right).
\label{eq_supp_gamma_rad_mapped}
\end{equation}
Similarly, the dissipative decay rate $\gamma_{\text{diss}}$ quantifies the rate of volumetric energy absorption due to the material's extinction coefficient $\kappa$. By mapping the spatial attenuation factor $\exp(-2k_0\kappa z)$ onto the temporal domain via the wave velocity, we obtain:
\begin{equation}
\gamma_{\text{diss}} \approx \frac{\omega_0 \kappa}{n}.
\label{eq_supp_gamma_diss_mapped}
\end{equation}

The total quality factor $Q$ of an open resonator is universally defined through its total energy loss rate as 
\begin{equation}
    Q = \frac{\omega_0}{2 (\gamma_{\text{rad}} + \gamma_{\text{diss}})}.
\end{equation}
On the perfect absorption trajectory governed by the critical coupling condition [Eq.~\eqref{eq_supp_critical_coupling}], we can substitute $\gamma_{\text{rad}} = \gamma_{\text{diss}}$, which simplifies the denominator to a purely dissipative term:
\begin{equation}
Q_{\text{cc}}(n) = \frac{\omega_0}{4 \gamma_{\text{diss}}}.
\label{eq_supp_Q_cc_step}
\end{equation}
Substituting the physical definition of $\gamma_{\text{diss}}$ from Eq.~\eqref{eq_supp_gamma_diss_mapped} into Eq.~\eqref{eq_supp_Q_cc_step} immediately provides the explicit functional dependence of the quality factor under critical coupling as a function of the layer's material parameters:
\begin{equation} 
Q_{\text{cc}}(n) = \frac{n}{4\kappa(n)}. 
\label{eq_supp_Q_cc_final} 
\end{equation}
Importantly, because the extinction coefficient $\kappa(n)$ is implicitly bound to the refractive index $n$ to satisfy the $r=0$ condition, $Q_{\text{cc}}$ exhibits highly nonlinear behavior. 

At the low-contrast operational point $n=1$, the exact analytical solution for perfect absorption yields an optimal thickness of $d = 3\lambda_0/8$ and an optimal extinction coefficient of $\kappa = \frac{2}{3\pi} W(3\pi) \approx 0.36$, where $W(x)$ represents the principal branch of the Lambert $W$ function. Substituting these values into Eq.~\eqref{eq_supp_Q_cc_final} yields the global sub-unity threshold of the quality factor:
\begin{equation}
Q_{\text{cc, min}} = \frac{1}{4 \cdot \frac{2}{3\pi} W(3\pi)} = \frac{3\pi}{8 W(3\pi)} \approx 0.693.
\label{eq_supp_Q_min_lambert}
\end{equation}
This establishes the exact mathematical proof for the critically damped sub-unity state ($Q < 1$) of the rapid-interference regime.

\section{Time-Domain Resonator Dynamics and Ultrafast Energy Dissipation}

To map the transient temporal evolution of the internal electromagnetic field, we examine the free decay of the resonator mode amplitude $a(t)$ following an initial excitation. Assuming no incident field blocks the channel ($s_{+}(t) = 0$ for $t > 0$), the core time-domain TCMT differential equation reads:
\begin{equation} 
\frac{da(t)}{dt} = (-i\omega_0 - \gamma_{\text{rad}} - \gamma_{\text{diss}})a(t). 
\label{eq_supp_diffeq} 
\end{equation}
Solving Eq.~\eqref{eq_supp_diffeq} with the initial condition $a(0) = a_0$ yields the time-dependent mode amplitude:
\begin{equation} 
a(t) = a_0 e^{-i\omega_0 t} e^{-(\gamma_{\text{rad}} + \gamma_{\text{diss}})t}. 
\label{eq_supp_amplitudesol}
\end{equation}

By invoking the definition of the total quality factor $Q = \frac{\omega_0}{2(\gamma_{\text{rad}} + \gamma_{\text{diss}})}$, we express the total energy decay rate as $\Gamma = \gamma_{\text{rad}} + \gamma_{\text{diss}} = \frac{\omega_0}{2Q}$. The normalized stored energy within the layer, defined as $\mathcal{E}(t) = |a(t)|^2 / |a_0|^2$, evolves as:
\begin{equation} 
\mathcal{E}(t) = \exp\left( -\frac{\omega_0 t}{Q} \right). 
\label{eq_supp_energysol} 
\end{equation}

To provide a universal comparison across different refractive index regimes, we normalize the time scale to the number of optical oscillation cycles $m = t/T$, where $T = 2\pi/\omega_0$ represents the optical period. Substituting $t = m \cdot \frac{2\pi}{\omega_0}$ into Eq.~\eqref{eq_supp_energysol} reveals the cycle-dependent energy envelope:
\begin{equation}
\mathcal{E}(m) = \exp\left( -\frac{2\pi m}{Q} \right). 
\label{eq_supp_cycle_energy} 
\end{equation}

Under the critical coupling state, $Q$ equals $Q_{\text{cc}}$. In the low-contrast mesoscopic regime ($n \approx 1$), the critical coupling quality factor reaches its absolute global minimum of $Q_{\text{cc, min}} \approx 0.69$ (the Butterworth limit). Evaluating the energy remaining in the cavity after exactly one full optical cycle ($m=1$) yields:
\begin{equation}
\mathcal{E}(1) = \exp\left( -\frac{2\pi \cdot 1}{0.69} \right) \approx \exp(-9.1) \approx 1.1 \times 10^{-4}.
\end{equation}
This indicates that the layer dampens the internal energy by approximately four orders of magnitude (${\sim}99.99\%$) within a single optical cycle, confirming a nearly instantaneous dissipation mechanism. 

For comparison, a classic high-contrast Dallenbach absorber operating at $n=10$ exhibits a significantly higher quality factor of $Q_{\text{cc}} \approx 4$. For the same single-cycle duration ($m=1$), the remaining energy envelope evaluates to:
\begin{equation}
\mathcal{E}(1) = \exp\left( -\frac{2\pi \cdot 1}{4} \right) \approx \exp(-1.57) \approx 0.21.
\end{equation}
Thus, the Dallenbach structure retains more than $20\%$ of its energy after one cycle, giving rise to multiwave temporal reflections and parasitic ringing artifacts in the time domain.

\end{document}